\documentclass{article}
\usepackage{arxiv}
\usepackage[alsoload=binary]{siunitx}
\usepackage{tabulary}
\usepackage{booktabs}
\usepackage{url}
\usepackage{amssymb,amsmath}
\usepackage{threeparttable}
\usepackage[shortcuts]{extdash}
\usepackage{multirow}
\usepackage{graphicx}
\usepackage{caption}
\usepackage{subcaption}
\usepackage{bbm}
\usepackage{natbib}

\DeclareMathOperator*{\argmax}{arg\,max}
\DeclareMathOperator*{\argmin}{arg\,min}

\title{Empirical investigation of state-of-the-art mean reversion strategies for equity markets}

	\author{Seung-Hyun Moon \\
	School of Computer Science \& Engineering \\ 
	Seoul National University \\ 
	1 Gwanak-ro, Gwanak-gu, Seoul 08826 \\
	Republic of Korea \\
	\texttt{shmoon@soar.snu.ac.kr} \\
	\And
	Yong-Hyuk Kim \\
	School of Software \\
	Kwangwoon University \\
	20 Kwangwoon-ro, Nowon-gu, Seoul 01897 \\
	Republic of Korea \\
	\texttt{yhdfly@kw.ac.kr}
	\And
	Byung-Ro Moon \\
	School of Computer Science \& Engineering \\ 
	Seoul National University \\ 
	1 Gwanak-ro, Gwanak-gu, Seoul 08826 \\
	Republic of Korea \\
	\texttt{moon@snu.ac.kr} \\
	}

\begin{document}
\maketitle




\begin{abstract}

Recent studies have shown that online portfolio selection strategies that exploit the mean reversion property can achieve excess return from equity markets. This paper empirically investigates the performance of state-of-the-art mean reversion strategies on real market data. The aims of the study are twofold. The first is to find out why the mean reversion strategies perform extremely well on well-known benchmark datasets, and the second is to test whether or not the mean reversion strategies work well on recent market data. The mean reversion strategies used in this study are the passive aggressive mean reversion (PAMR) strategy, the on-line moving average reversion (OLMAR) strategy, and the transaction cost optimization (TCO) strategies. To test the strategies, we use the historical prices of the stocks that constitute S\&P 500 index over the period from 2000 to 2017 as well as well-known benchmark datasets. Our findings are that the well-known benchmark datasets favor mean reversion strategies, and mean reversion strategies may fail even in favorable market conditions, especially when there exist explicit or implicit transaction costs.

\end{abstract}

\keywords{Online portfolio selection, Rebalancing investments, Mean reversion.}


\section{Introduction}


Mean reversion is a tendency to move back to the average. It implies that a stock with a recent above-average return is subject to produce a below-average return during the next period, and vice versa. Whether or not such a tendency exists in stock price movements has been controversial in recent decades. \cite{10.2307/1833108} find negative autocorrelations of stock returns, \cite{Balvers00meanreversion} discover strong evidence of mean reversion in national equity indexes, and \cite{Gropp2004537} finds evidences of mean reversion in portfolio prices. However, \cite{Kim01051991} find no evidence of mean reversion after the Second World War, and \cite{GeoffreyBooth201668} show that there is no momentum-reversal anomaly in the U.\@S.\@ stock returns from 1962 to 2013.


Meanwhile, there have been a significant improvement in online portfolio selection. The online portfolio selection involves selling securities that are expected to fall in price and buying securities that are expected to rise within a portfolio. Only past price information can be used to determine the proportion of assets in the portfolio. Although the framework of well-performing strategies can be used to predict the price movements of multiple assets~\citep{Roch2013Histogram}, the natural objective of online portfolio selection is to maximize the cumulative wealth of an investor. Refer to \cite{Li2014Online} for a comprehensive survey on online portfolio selection.

The most widely used dataset for online portfolio selection is the NYSE(O) dataset, which includes the historical daily prices of 36 stocks listed in the New York Stock Exchange (NYSE) for 22 years. Various strategies have been proposed and tested using the dataset~\citep{Agarwal:2006:APM:1143844.1143846,Borodin:2004:WLB:1622467.1622484,MAFI:MAFI274,MAFI:MAFI058,Huang:2018:CFR:3210369.3200692,Levina2008Portfolio,PAMR,Li:2013:CWM:2435209.2435213,Li2015104,doi:10.1080/14697688.2017.1357831,Singer1997Switching}. However, it is worthwhile to test the strategies with as many datasets as possible to see how well they perform on unseen data. 


Recently, three kinds of mean reversion strategies that outperform other state-of-the-art strategies have been developed using machine learning techniques. One is the passive aggressive mean reversion (PAMR) strategy~\citep{PAMR}, another is the online moving average reversion (OLMAR) strategy~\citep{Li2015104}, and the other is the transaction cost optimization (TCO) strategies~\citep{doi:10.1080/14697688.2017.1357831}. The PAMR exploits the single-period mean reversion property and the OLMAR exploits the multi-period mean reversion property. The TCO can exploit the single-period or the multi-period mean reversion property depending on the settings. While a simple buy-and-hold strategy increases its wealth by 14 times, the three strategies increase the wealth by over ten trillion times on the NYSE(O) dataset.  

In this paper, we first attempt to find out why mean reversion strategies work extremely well on well-known benchmark datasets. For this aim, we devise simple mean reversion strategies that behave similarly to the state-of-the-art mean reversion strategies. The strategies we devised are much easier to comprehend than their counterparts so that we can realize the property of the datasets on which mean reversion strategies perform well. The performance of the simple strategies on the benchmark datasets reveal the characteristic feature of the datasets.

On well-known benchmark datasets, \cite{OLPS15} report that OLMAR can outperform the other strategies with a high confidence, and \cite{doi:10.1080/14697688.2017.1357831} state that TCO can effectively handle reasonable transaction costs. However, it is wondered whether or not the strategies can be successful on other datasets. If they work well on most datasets, it is a threat to the efficient-market hypothesis, which implies that it is impossible to beat the market. Thus, we test whether or not the mean reversion strategies work well on recent market data. Our empirical results show the potential pitfalls of the state-of-the-art mean reversion strategies.

\section{Framework}\label{sec:framework}
Suppose that we have $m$ stocks to trade for $n$ days. The closing prices of the stocks on the $t^\mathrm{th}$ day is denoted by $\mathbf{p}_t=(p_t(1), p_t(2), ..., p_t(m)) \in \mathbb{R}_+^m$, where $p_t(j)$ is the closing price of the $j^\mathrm{th}$ stock, and the daily return of the stocks on the $t^\mathrm{th}$ day is denoted by a price relative vector, $\mathbf{x}_t=(x_t(1), x_t(2), ..., x_t(m)) \in \mathbb{R}_+^m$, where $x_t(j)=p_t(j)/p_{t-1}(j)$. At the beginning of the $t^\mathrm{th}$ day, we specify a portfolio vector, $\mathbf{b}_t=(b_t(1), b_t(2), ..., b_t(m)) \in \mathbb{R}_+^m$, where $b_t(j)$ is the proportion of the $j^\mathrm{th}$ stock to our wealth and $\|\mathbf{b}_t\|=1$. After $\mathbf{x}_t$ is revealed, the daily return on the $t^\mathrm{th}$ day is $\mathbf{b}_t \cdot \mathbf{x}_t$, and compound return for $n$ days is $\prod_{t=1}^n \mathbf{b}_t \cdot \mathbf{x}_t$. The aim of online portfolio selection is to maximize the compound return. It is assumed that all stocks are arbitrarily divisible and we can trade them at their last closing prices. 


Let $\mathrm{BAH}_{\mathbf{b}}$ denote a buy-and-hold (BAH) strategy, which allocates investor's initial wealth according to the portfolio vector $\mathbf{b}$ on the first trading day and does not make any transactions until the last trading day. When the portfolio vector of BAH is uniform, i.e., $\mathbf{b}=(1/m,...,1/m)$, the BAH strategy is denoted by $\mathrm{BAH}_{\mathbf{U}}$. Let $\mathrm{CRP}_{\mathbf{b}}$ denote a constant rebalanced portfolio (CRP), which reallocates investor's wealth according to the portfolio vector $\mathbf{b}$ at the beginning of every trading day. When the portfolio vector of CRP is uniform, i.e., $\mathbf{b}=(1/m,...,1/m)$, the CRP is denoted by $\mathrm{CRP}_{\mathbf{U}}$. In spite of its simplicity, $\mathrm{CRP}_{\mathbf{U}}$ often outperforms $\mathrm{BAH}_\mathbf{U}$ and other more sophisticated rebalancing strategies such as \emph{universal portfolios}~\citep{Agarwal:2006:APM:1143844.1143846,Borodin:2004:WLB:1622467.1622484,MAFI:MAFI1}. The $\mathrm{CRP}_\mathbf{U}$ exploits the mean reversion property of stock market since it sells stocks that performed above average and buys stocks that performed below average on the previous trading day. The performances of $\mathrm{BAH}_\mathbf{U}$ and $\mathrm{CRP}_\mathbf{U}$ provide benchmarks for this study.

The PAMR is a mean reversion strategy that exploits the single-period mean reversion. The PAMR utilizes the passive aggressive online learning algorithm~\citep{Crammer:2006:OPA:1248547.1248566} to find the mean revertible portfolio vector that is as close as possible to the current vector, i.e., 
\begin{align}
	\begin{split}
		\mathbf{b}_{t+1}=\,& \argmin_{\mathbf{b} \in \mathbb{R}_+^m} \frac{1}{2}\|\mathbf{b}-\mathbf{b}_{t}\|^2 \\
		\text{subject to }& \, \mathbf{b} \cdot \mathbf{x}_{t} \leq \epsilon \ \textrm{and} \ \|\mathbf{b}\|=1,
\end{split}
\end{align}
where $\epsilon \in \mathbb{R_+}$ is a user-defined parameter. We set $\epsilon$ to 0.5 which is the default value in \cite{PAMR}.

The PAMR is a single-period mean reversion strategy since its portfolio vector depends on $\mathbf{x}_{t}$, the latest price relative vector. The strategy finds a portfolio vector whose latest return is confined to $\epsilon$ in order to benefit from the single-period reversion ($\mathbf{b} \cdot \mathbf{x}_{t} \leq \epsilon$). To better understand why the strategy works well, we devise a simple mean reversion (SMR) strategy that exploits the single-period mean reversion in the simplest way. The SMR puts all the money into the worst performer of the previous day. If there are multiple stocks that satisfy the investment criterion, we evenly buy all the stocks that satisfy the criterion, i.e.,
\begin{align}
	\begin{split}
	\mathbf{b}_{t+1}=\argmin_{\mathbf{b} \in \mathbb{R}_+^m}& \mathbf{b} \cdot \mathbf{x}_{t} \\
		\text{subject to} \sum\limits_{i=1}^{m} \sum\limits_{j=1}^{m} |b(i) - b(j)|\mathbb{I}[x_t(i)& = x_t(j)] = 0 \ \textrm{and} \ \|\mathbf{b}\|=1,
	\end{split}
\end{align}
where $\mathbb{I}[\cdot]$ is the indicator function.

Another state-of-the-art mean reversion strategy is OLMAR which exploits the multi-period mean reversion property. Instead of using the latest price relative vector, the OLMAR employs a simple moving average (SMA) to determine a portfolio vector. The simple moving average is the unweighted mean of the closing prices. The $\mathbf{SMA}_{t}(w) \in \mathbb{R}_+^m$ represents the SMA on the $t^\mathrm{th}$ day, i.e.,
\begin{align}
	\begin{split}
		\mathbf{SMA}_{t}(w)=\frac{1}{w}\sum\limits_{i=t-w+1}^{t}{\mathbf{p}_i},
\end{split}
\end{align} 
where $w \in \mathbb{Z}_+$ is the window size. The OLMAR finds the mean revertible portfolio vector that is as close as possible to the current vector, and thus has a similar formulation to PAMR, i.e.,
\begin{align}
	\begin{split}
		\mathbf{b}_{t+1}=&\argmin_{\mathbf{b} \in \mathbb{R}_+^m} \frac{1}{2}\|\mathbf{b}-\mathbf{b}_{t}\|^2 \\
		\text{subject to }& \mathbf{b} \cdot \tilde{\mathbf{x}}_{t+1} \geq \epsilon\ \textrm{and} \ \|\mathbf{b}\|=1,
\end{split}
\end{align}
where $\epsilon \in \mathbb{R_+}$ is a user-defined parameter and $\tilde{\mathbf{x}}_{t+1} \in \mathbb{R}_+^m$ is the predicted price relative which can be calculated when $\mathbf{p}_{t}$ is revealed. The OLMAR uses the SMA to calculate the predicted price relative, i.e., 
\begin{align}
	\begin{split}
		\tilde{\mathbf{x}}_{t+1}(w) = \mathbf{SMA}_{t}(w)\oslash\mathbf{p}_{t},
	\end{split}
\end{align}
where $\oslash$ denotes an element-wise division. We set $\epsilon$ to 10 and $w$ to 5, which are the default values used in \cite{Li2015104}.

Under the assumption of the multi-period mean reversion, current price tends to revert to the mean. Thus, the value of the predicted price relative represents the likelihood of gaining a profit. The higher the value of the predicted price relative, the higher the chance of profit. To test whether the assumption of the OLMAR holds, we devise a simple moving average reversion (SMAR) strategy, which invests into the stock that has the highest predicted price relative. If there are multiple stocks that satisfy the criterion, we evenly buy all the stocks that satisfy the criterion, i.e.,
\begin{align}
	\begin{split}
		\mathbf{b}_{t+1}=\argmax_{\mathbf{b} \in \mathbb{R}_+^m}& \mathbf{b} \cdot \mathbf{\tilde{x}}_{t+1} \\
		\text{subject to} \sum\limits_{i=1}^{m} \sum\limits_{j=1}^{m} |b(i) - b(j)|\mathbb{I}[\tilde{x}_{t+1}(i)& = \tilde{x}_{t+1}(j)] = 0 \ \textrm{and} \ \|\mathbf{b}\|=1.
	\end{split}
\end{align}
The SMAR uses the SMA to calculate $\mathbf{\tilde{x}}_{t+1}$ in the same way as the OLMAR.

The other state-of-the-art mean reversion strategy is TCO, which is devised to perform well in the case of non-zero transaction costs. Since insignificant trades can incur unnecessary transaction costs, the TCO only accepts changes larger than a certain threshold in the portfolio vector. Let $\hat{\mathbf{b}_{t}} \in \mathbb{R}_+^m$ be the last price adjusted portfolio vector and $\tilde{\mathbf{b}}_{t+\frac{1}{2}} \in \mathbb{R}^m$ be the weights to be adjusted after the $t^\mathrm{th}$ trading day, then the TCO uses the predicted price relative to determine $\tilde{\mathbf{b}}_{t+\frac{1}{2}}$, i.e., 
\begin{align}
	\tilde{\mathbf{b}}_{t+\frac{1}{2}}=\eta\left(\frac{\tilde{\mathbf{x}}_{t+1}}{\hat{\mathbf{b}_{t}}\cdot\tilde{\mathbf{x}}_{t+1}}-\frac{1}{m}\mathbf{1}\cdot\frac{\tilde{\mathbf{x}}_{t+1}}{\hat{\mathbf{b}_{t}}\cdot\mathbf{\tilde{x}}_{t+1}}\right),
\end{align}
where $\hat{\mathbf{b}}_{t}= (\mathbf{b}_{t} \odot \mathbf{x}_{t}) / (\mathbf{b}_{t} \cdot \mathbf{x}_{t})$, $\odot$ is an element-wise product, and $\eta \in \mathbb{R}_+$ is a user-defined parameter. The negative components of $\tilde{\mathbf{b}}_{t+\frac{1}{2}}$ have lower predicted price relatives than average, so their weights will be decreased on the next trading day, and vice versa, i.e.,
\begin{align}
	b_{t+1}(j) = \hat{b}_{t}(j) + \left(\mathbb{I}\left[\tilde{b}_{t+\frac{1}{2}}(j) \geq 0\right] - \mathbb{I}\left[\tilde{b}_{t+\frac{1}{2}}(j) < 0\right]\right)\cdot\max\left(\left|\tilde{b}_{t+\frac{1}{2}}(j)\right|-\lambda, 0\right),
\end{align}
for all $1 \leq j \leq m$, where $\lambda \in \mathbb{R}_+$ is a user-defined parameter. Only changes greater than the threshold of $\lambda$ are applied to the next portfolio vector. We set $\eta$ to 10 and $\lambda$ to $10 \times \eta \times \gamma$, where $\gamma$ denotes the rate of transaction costs.\footnote{\cite{doi:10.1080/14697688.2017.1357831} states that they set $\lambda$ to $10 \times \gamma$. However, their implementation sets $\lambda$ to $10 \times \eta \times \gamma$, which performs better than $10 \times \gamma$ and coincides with the result of their study.} Since $\mathbf{b}_{t+1}$ may not satisfy the constraints of the portfolio vector, $\mathbf{b}_{t+1}$ is normalized, i.e.,
\begin{align}
	\begin{split}
	\mathbf{b}_{t+1}&=\, \argmin_{\mathbf{b} \in \mathbb{R}_+^m} \|\mathbf{b}-\mathbf{b}_{t+1}\|^2 \\
		&\text{subject to } \, \|\mathbf{b}\|=1.
	\end{split}
\end{align}

There are two versions of the TCO depending on how the predictive price relatives are calculated. The TCO-1 uses the single-period mean reversion property, and the TCO-2 uses the multi-period mean reversion property, i.e.,
\begin{align}
	\tilde{\mathbf{x}}_{t+1}=\begin{cases}
		\mathbf{1} \oslash \mathbf{x}_{t}   		& \text{in TCO-1}\\
		\mathbf{SMA}_{t}(5)	\oslash \mathbf{x}_{t}	& \text{in TCO-2.}\\
	\end{cases}
\end{align}

\section{Data}\label{sec:data}
We use two sets of historical daily data to test mean reversion strategies. The first set of the data is summarized in Table~\ref{tbl:dataset1}. The name of each data represents the universe of the data, which is a stock market index or a stock exchange. The first set consists of the data that have been widely used by other researchers. Refer to \cite{PAMR} for the details of the data on the first set.

\begin{table}\footnotesize
\caption{Summary of the first set of historical daily data.}
\label{tbl:dataset1}
\centering
\begin{threeparttable}
\begin{tabulary}{\linewidth}{Lcccccc}
\toprule[\heavyrulewidth]
Name		&Period		&Days	&Assets\tnote{a}	&Max($x_t(j)$)\tnote{b}	&Min($x_t(j)$)\tnote{c}\\
\midrule[\lightrulewidth]
NYSE(O)				&1962 - 1984	&5651	&36	&1.3529		&0.7500\\	
NYSE(N)				&1985 - 2010	&6431	&23	&1.8146		&0.4545\\	
DJIA				&2001 - 2003	&507	&30	&1.2012		&0.4027\\
SP500				&1998 - 2003	&1276	&25	&1.2439		&0.6976\\
TSE				&1994 - 1998	&1259	&88	&1.9392		&0.3685\\
MSCI				&2006 - 2010	&1043	&24	&1.1663		&0.8274\\
\bottomrule[\heavyrulewidth]
\end{tabulary}
\begin{tablenotes}
\item[a] The number of assets in the portfolio.
\item[a] The highest daily return of a single asset on the dataset.
\item[b] The lowest daily return of a single asset on the dataset.
\end{tablenotes}
\end{threeparttable}
\end{table}

	For the second set of the data, we collect the historical prices of 505 stocks that were the components of the S\&P 500 on October 31, 2018. We obtain the data from Yahoo Finance. Out of the 505 stocks, we exclude the 111 stocks that were not listed on January 1, 2000 and 5 stocks whose data could not be retrieved correctly. We then sort the remaining 389 stocks in alphabetical order by the ticker symbols of the companies and make 10 portfolios each of which contains 38 or 39 stocks in the order of the sort. The second set of the data is summarized in Table~\ref{tbl:dataset2}.\footnote{The second set of the data in csv and mat format will be available at the site: \url{https://github.com/uramoon/mr_dataset}.}

\begin{table}\footnotesize
\caption{Summary of the second set of historical daily data.}
\label{tbl:dataset2}
\centering
\begin{threeparttable}
\begin{tabulary}{\linewidth}{Lcrccc}
\toprule[\heavyrulewidth]
Name			&Period		&Days	&Assets\tnote{a}	&Max($x_t(j)$)\tnote{b}	&Min($x_t(j)$)\tnote{c}\\
\midrule[\lightrulewidth]
SP500(0)				&2000 - 2017	&4527	&39	&1.6600		&0.3921\\
SP500(1)				&2000 - 2017	&4527	&39	&1.5782		&0.5494\\
SP500(2)				&2000 - 2017	&4527	&39	&1.6296		&0.5402\\
SP500(3)				&2000 - 2017	&4527	&39	&1.6037		&0.4133\\
SP500(4)				&2000 - 2017	&4527	&39	&2.0236		&0.4844\\
SP500(5)				&2000 - 2017	&4527	&39	&1.5425		&0.3948\\
SP500(6)				&2000 - 2017	&4527	&39	&1.8698		&0.5873\\
SP500(7)				&2000 - 2017	&4527	&39	&1.7545		&0.3195\\
SP500(8)				&2000 - 2017	&4527	&39	&1.5605		&0.4096\\
	SP500(9)				&2000 - 2017	&4527	&38\tnote{d}	&2.0101		&0.3811\\
\bottomrule[\heavyrulewidth]
\end{tabulary}
\begin{tablenotes}
\item[a] The number of assets in the portfolio.
\item[a] The highest daily return of a single asset on the dataset.
\item[b] The lowest daily return of a single asset on the dataset.
\item[d] One less stocks than the others since the total number of stocks is 389. 
\end{tablenotes}
\end{threeparttable}
\end{table}

\section{Empirical results}\label{sec:empirical}
Experiments are conducted to investigate the performance of the strategies described in this paper. We implement SMR and SMAR in MATLAB and use the implementations of \cite{Li2015104,JMLR:v17:15-317,doi:10.1080/14697688.2017.1357831} for the rest of the strategies. As stated earlier, $\mathrm{BAH}_\mathbf{U}$ and $\mathrm{CRP}_\mathbf{U}$ provide benchmarks for rebalancing strategies since they can be easily applied in the real stock market.

In actual trading, traders have to pay transaction costs when they trade. Commissions and taxes are explicit transaction costs, while bid-ask spreads and price-impact costs are implicit transaction costs. \cite{10.2307/4480093} find that explicit costs are at least \SI{0.13}{\percent} and implicit costs are at least \SI{0.11}{\percent} on institutional trades data. The lowest total costs are \SI{0.26}{\percent} for seller-initiated trades, and \SI{0.31}{\percent} for buyer-initiated trades in NYSE and NYSE American. To incorporate all expenses that can be incurred in actual trading, transaction costs are taken into account in the experiments. We adopt the \emph{proportional commission} model used by \citet{Borodin:2004:WLB:1622467.1622484} and \citet{PAMR,Li:2013:CWM:2435209.2435213,Li2015104}. Given a rate of transaction costs $\gamma \in [0, 1]$, the model incurs transaction costs at a rate of $\gamma$ for each buy and for each sell so that the daily return on the $t^\mathrm{th}$ day becomes $(\mathbf{b}_t \cdot \mathbf{x}_t) \times (1 - \gamma \cdot \|\mathbf{b}_{t} - \mathbf{b}_{t-1}\|$). Since transaction costs depend on trading volume, strategies with higher volatility suffer more from transaction costs. 

Table~\ref{tbl:cw1} shows the cumulative wealth of each strategy using the first set of data. The initial wealth of each strategy is 1, and transaction costs are not considered. The results of SMR and SMAR are newly obtained ones, and the results of the other strategies are consistent with their original papers. In Table~\ref{tbl:cw1}, the results of $\mathrm{BAH}_{\mathbf{U}}$ indicates that the first set of data includes both upward and downward trends. All of the mean reversion strategies work better than $\mathrm{BAH}_\mathbf{U}$ and $\mathrm{CRP}_\mathbf{U}$ on all dataset except DJIA; i.e., the SMR and PAMR fail to outperform $\mathrm{BAH}_{\mathbf{U}}$ on DJIA. Still, the overall performances of mean reversion strategies are very impressive, and it is notable that in all datasets except TSE, the top two strategies are based on the multi-period mean reversion property. It is also noteworthy that very simple mean reversion strategies we devise (SMR and SMAR) produce the best results on five datasets. The success of the very simple strategies shows that there are exceptionally strong mean reversion tendencies in the well-known benchmark dataset, which might have caused mean reversion strategies to perform well.

\sisetup{output-exponent-marker = \ensuremath{\mathrm{e}}}
\begin{table*}\footnotesize
\caption{Cumulative wealth achieved by various strategies using the first set of historical daily data. Transaction costs are not considered. The top two results are indicated in bold type for each dataset.}
\label{tbl:cw1}
\centering
\begin{threeparttable}
\begin{tabulary}{\hsize}{Lrrrrrrrr}
\toprule[\heavyrulewidth]
	Data	&$\mathrm{BAH}_{\mathbf{U}}$	&$\mathrm{CRP}_{\mathbf{U}}$	&SMR				&SMAR			&PAMR		&OLMAR		&TCO-1		&TCO-2\\
\midrule[\lightrulewidth]
	NYSE(O)	&14.50							&27.08							&4.39e+15			&\textbf{5.30e+16}		&5.14e+15	&\textbf{3.68e+16}	&1.35e+14	&1.40e+13\\
	NYSE(N)	&18.06							&31.55							&7.15e+05			&\textbf{3.66e+08}		&1.25e+06	&\textbf{2.54e+08}	&9.15E+06	&2.43e+07\\
	DJIA	&0.76							&0.81							&0.59				&\textbf{2.54}	&0.68		&2.06		&1.55		&\textbf{2.16}\\
	SP500	&1.34							&1.65							&8.88				&\textbf{16.99}	&5.09		&5.83		&5.33		&\textbf{10.70}\\
	TSE		&1.61							&1.60							&\textbf{1.44e+03}	&84.66			&264.86		&\textbf{424.80}		&149.00		&153.05\\
	MSCI	&0.91							&0.93							&10.23				&\textbf{13.99}			&15.23		&\textbf{16.39}		&9.68		&5.68\\
\bottomrule[\heavyrulewidth]
\end{tabulary}
\end{threeparttable}
\end{table*}

In \cite{doi:10.1080/14697688.2017.1357831}, the state-of-the-art mean reversion strategies (PAMR, OLMAR, and TCOs) are tested under the rates of transaction costs at \SI{0.25}{\percent} and \SI{0.5}{\percent}, and it is claimed that TCOs are able to withstand reasonable transaction costs. Table~\ref{tbl:cw1tc} shows the cumulative wealth when transaction costs are set to \SI{0.25}{\percent} in the proportional commission model. We can see that the results of strategies that do not consider transaction costs in determining portfolio vectors are very poor on all datasets except NYSE(O) and TSE. However, the performances of TCOs are very promising even in the presence of transaction costs; i.e., the TCOs produce the top two results on five datasets. 



\begin{table*}\footnotesize
	\caption{Cumulative wealth achieved by various strategies using the first set of historical daily data. Transaction costs are set to \SI{0.25}{\percent} in the proportional commission model. The top two results are indicated in bold type for each dataset.}
\label{tbl:cw1tc}
\centering
\begin{threeparttable}
\begin{tabulary}{\hsize}{Lrrrrrrrrrr}
\toprule[\heavyrulewidth]
	Data	&$\mathrm{BAH}_{\mathbf{U}}$	&$\mathrm{CRP}_{\mathbf{U}}$	&SMR				&SMAR			&PAMR		&OLMAR		&TCO-1				&TCO-2\\
\midrule[\lightrulewidth]
	NYSE(O)	&14.46							&22.93							&1.47e+04			&1.62e+08		&2.09e+05	&2.98e+07	&\textbf{5.53e+09}	&\textbf{3.86e+07}\\
	NYSE(N)	&18.01							&25.92							&0.00				&0.47			&0.00		&0.05		&\textbf{3.79e+03}	&\textbf{2.00e+03}\\
	DJIA	&0.76							&0.80							&0.05				&0.47			&0.09		&0.33		&\textbf{0.83}		&\textbf{1.19}\\
	SP500	&1.34							&\textbf{1.57}					&0.02				&0.24			&0.03		&0.07		&0.60				&\textbf{2.16}\\
	TSE		&1.61							&1.52							&3.74				&1.12			&2.11		&4.29		&\textbf{7.73}		&\textbf{31.54}\\
	MSCI	&0.90							&0.90							&0.08				&0.40			&0.15		&0.34		&\textbf{1.52}		&\textbf{1.42}\\
\bottomrule[\heavyrulewidth]
\end{tabulary}
\end{threeparttable}
\end{table*}

To test if mean reversion strategies perform well on various and recent datasets, experiments are conducted on the second set of data. Table~\ref{tbl:cw2} shows the cumulative wealth achieved by each strategy. The result of $\mathrm{BAH}_{\mathbf{U}}$ shows that every dataset on the second set has an upward trend in the long term. The $\mathrm{BAH}_{\mathbf{U}}$ increases the wealth by over six times on every dataset. A sensible strategy is expected to profit from these datasets. However, all the mean reversion strategies fail to surpass $\mathrm{BAH}_{\mathbf{U}}$ and $\mathrm{CRP}_{\mathbf{U}}$ on SP500(0) and SP500(6). Three strategies (SMR, SMAR, and OLMAR) even suffer a serious loss on SP500(0) while $\mathrm{BAH}_{\mathbf{U}}$ increases its wealth by over nine times. Nevertheless, the mean reversion strategies perform well on many datasets. The multi-period mean reversion strategies generally work better than the single-period counterparts. It is noteworthy that simple strategies work as well as the state-of-the-art mean reversion strategies on the second set of data; the SMAR produces the best results on three datasets.


\sisetup{output-exponent-marker = \ensuremath{\mathrm{e}}}
\begin{table*}\footnotesize
\caption{Cumulative wealth achieved by various strategies using the second set of historical daily data. Transaction costs are not considered. The top two results are indicated in bold type for each dataset.}
\label{tbl:cw2}
\centering
\begin{threeparttable}
\begin{tabulary}{\hsize}{Lrrrrrrrrrr}
\toprule[\heavyrulewidth]
	Data	&$\mathrm{BAH}_{\mathbf{U}}$	&$\mathrm{CRP}_{\mathbf{U}}$	&SMR			&SMAR				&PAMR			&OLMAR				&TCO-1			&TCO-2\\
\midrule[\lightrulewidth]
	SP500(0)&\textbf{9.44}					&\textbf{13.36}					&0.43			&0.14				&1.52			&0.11				&2.14			&2.29\\
	SP500(1)&8.31							&\textbf{9.84}					&0.42			&3.12				&5.45			&1.79				&\textbf{11.68}	&4.33\\
	SP500(2)&6.97							&8.50							&130.11			&190.11				&198.24			&\textbf{270.28}	&232.07			&\textbf{353.57}\\
	SP500(3)&6.68							&9.09							&47.01			&\textbf{3.07e+03}	&103.52			&\textbf{839.93}	&230.16			&189.85\\
	SP500(4)&7.54							&8.73							&36.59			&34.76				&\textbf{95.58}	&23.50				&\textbf{93.28}	&84.79\\
	SP500(5)&6.39							&7.82							&49.95			&\textbf{334.40}	&162.52			&42.12				&191.22			&\textbf{316.48}\\
	SP500(6)&\textbf{24.29}					&11.08							&0.97			&6.74				&9.48			&3.57				&\textbf{12.80}	&9.93\\
	SP500(7)&7.92							&11.88							&4.98			&13.61				&16.24			&31.23				&\textbf{40.62}	&\textbf{73.39}\\
	SP500(8)&8.90							&12.38							&16.29			&\textbf{105.91}	&\textbf{154.65}&59.30				&88.85			&22.59\\
	SP500(9)&6.49							&9.17							&78.2			&\textbf{104.34}	&101.82			&16.35				&52.32			&\textbf{155.14}\\
\bottomrule[\heavyrulewidth]
\end{tabulary}
\end{threeparttable}
\end{table*}

Mean reversion strategies are tested on the second set of data in the presence of transaction costs. Table~\ref{tbl:cw2tc} shows the cumulative wealth when transaction costs are set to \SI{0.25}{\percent}. All the mean reversion strategies, excluding TCOs, spend all assets due to transaction costs. Even TCOs, strategies that takes transaction costs into account, fail to get excess returns from most datasets. We can see that applying mean reversion strategies to actual investments based on the result of well-known benchmark datasets can be very dangerous especially when there exist transaction costs. 

\sisetup{output-exponent-marker = \ensuremath{\mathrm{e}}}
\begin{table*}\footnotesize
\caption{Cumulative wealth achieved by various strategies using the second set of historical daily data. Transaction costs are set to \SI{0.25}{\percent} in the proportional commission model. The top two results are indicated in bold type for each dataset.}
\label{tbl:cw2tc}
\centering
\begin{threeparttable}
\begin{tabulary}{\hsize}{Lrrrrrrrrrr}
\toprule[\heavyrulewidth]
	Data	&$\mathrm{BAH}_{\mathbf{U}}$	&$\mathrm{CRP}_{\mathbf{U}}$	&SMR			&SMAR				&PAMR			&OLMAR				&TCO-1			&TCO-2\\
\midrule[\lightrulewidth]
	SP500(0)&\textbf{9.41}					&\textbf{11.47}					&0.00			&0.00				&0.00			&0.00				&0.01			&0.04\\
	SP500(1)&\textbf{8.29}					&\textbf{8.59}					&0.00			&0.00				&0.00			&0.00				&0.05			&0.25\\
	SP500(2)&\textbf{6.95}					&\textbf{7.46}					&0.00			&0.00				&0.00			&0.00				&1.07			&5.19\\
	SP500(3)&\textbf{6.66}					&\textbf{7.93}					&0.00			&0.00				&0.00			&0.00				&0.88			&4.98\\
	SP500(4)&\textbf{7.52}					&\textbf{7.57}					&0.00			&0.00				&0.00			&0.00				&0.43			&1.55\\
	SP500(5)&6.37					&\textbf{6.82}					&0.00			&0.00				&0.00			&0.00				&0.45			&\textbf{7.22}\\
	SP500(6)&\textbf{24.23}					&\textbf{9.54}					&0.00			&0.00				&0.00			&0.00				&0.07			&0.62\\
	SP500(7)&\textbf{7.90}					&\textbf{10.35}					&0.00			&0.00				&0.00			&0.00				&0.15			&1.39\\
	SP500(8)&\textbf{8.88}					&\textbf{10.84}					&0.00			&0.00				&0.00			&0.00				&0.09			&0.19\\
	SP500(9)&6.47					&\textbf{8.00}					&0.00			&0.00				&0.00			&0.00				&0.17			&\textbf{13.10}\\
\bottomrule[\heavyrulewidth]
\end{tabulary}
\end{threeparttable}
\end{table*}

\section{Conclusion}\label{sec:conclusion}
We investigate the performance of mean reversion strategies on various datasets. Unlike the other mean reversion strategies used in this work, $\mathrm{CRP}_\mathbf{U}$ generally shows consistent improvement over $\mathrm{BAH}_\mathbf{U}$ regardless of transaction costs. Although, on the other hand, the others overall work well without transaction costs, they show poor results with transaction costs especially on the datasets we collected. Given a portfolio, PAMR and OLMAR should be used with caution unless there is a good reason to believe that their counterparts (SMR and SMAR) will also perform well on the portfolio. Our empirical results also imply that extra caution should be taken in using the strategies in the case that there exist transaction costs. 


On the first set of data, it is very surprising that SMR and SMAR, very simple mean reversion strategies which we devise for this study, perform as well as more sophisticated mean reversion strategies. We also discover that if we invested our wealth to the company ``kin\_ark'' in NYSE datasets only when its price had fallen in the previous day, our wealth would have been multiplied by 2.15e+9 in the NYSE(O) and 1.90e+4 in the NYSE(N) at the end of the period. The BAH of the single asset ``kin\_ark'', however, increases initial wealth $2.05$ times in NYSE(O) and $4.13$ times in NYSE(N). Thus, we conclude that the datasets NYSE(O) and NYSE(N) do not have representative nature to be standard benchmarks. They are biased to the mean reversion characterstics. The second set of data, which we collect, can be alternative benchmark datasets for online portfolio selection to develop more robust rebalancing strategies.


\section*{Acknowledgments}
We are grateful to Dr.\@ Seung-Kyu Lee for his helpful comments.

\section*{References}
\bibliographystyle{apa}
\bibliography{mr}
\end{document}